\documentclass[preprint, amsmath,amssymb, nobibnotes, aps, prb, superscriptaddress, citeautoscript]{revtex4-2}

\usepackage{graphicx}
\usepackage{dcolumn}
\usepackage{bm}
\usepackage{hyperref}
\usepackage[mathlines]{lineno}
\usepackage{braket}
\DeclareMathAlphabet\mathbfcal{OMS}{cmsy}{b}{n}

\begin{document}

\title{Automated all-functionals infrared and Raman spectra}

\author{Lorenzo Bastonero}
\email{lbastone@uni-bremen.de}
\affiliation{U Bremen Excellence Chair,
Bremen Center for Computational Materials Science,
and MAPEX Center for Materials and Processes,
University of Bremen,
D-28359 Bremen, 
Germany}
\author{Nicola Marzari}
\affiliation{U Bremen Excellence Chair,
Bremen Center for Computational Materials Science,
and MAPEX Center for Materials and Processes,
University of Bremen,
D-28359 Bremen, 
Germany}
\affiliation{Theory and Simulation of Materials (THEOS), and National Centre for Computational Design and Discovery of Novel Materials (MARVEL), \'Ecole Polytechnique F\'ed\'erale de Lausanne (EPFL), CH-1015 Lausanne, Switzerland}
\affiliation{Laboratory for Materials Simulations, Paul Scherrer Institut, 5232 Villigen PSI, Switzerland}

\begin{abstract}
Infrared and Raman spectroscopies are ubiquitous techniques employed in many experimental laboratories, thanks to their fast and non-destructive nature able to capture materials' features as spectroscopic fingerprints. Nevertheless, these measurements frequently need theoretical support in order to unambiguously decipher and assign complex spectra. Linear-response theory provides an effective way to obtain the higher-order derivatives needed, but its applicability to modern exchange-correlation functionals remains limited. Here, we devise an automated, open-source, user-friendly approach based on ground-state density-functional theory and the electric enthalpy functional to allow seamless calculations of first-principles infrared and Raman spectra. By employing a finite-displacement and finite-field approach, we allow for the use of any functional, as well as an efficient treatment of large low-symmetry structures. Additionally, we propose a simple scheme for efficiently sampling the Brillouin zone with different electric fields. To demonstrate the capabilities of our approach, we provide illustrations using the ferroelectric LiNbO$_3$ crystal as a paradigmatic example. We predict infrared and Raman spectra using various (semi)local, Hubbard corrected, and hybrid functionals. Our results also show how PBE0 and extended Hubbard functionals yield in this case the best match in term of peak positions and intensities, respectively.
\end{abstract}

\date{\today}
\maketitle

\section*{Introduction}
Vibrational spectroscopies are one of the most powerful, fast and reliable methods for materials'
structure identification and characterization. 
Among the different techniques probing atomic motions,
infrared (IR) absorption and Raman experiments are 
widespread in most experimental set-ups, thanks to 
their low cost and precision in identifying microscopic features.
Despite these advantages, data interpretation for many
cases calls for theoretical support.
Density-functional theory (DFT) is most often employed 
as a reliable framework for accurate predictions of such spectra,
providing in principle exact estimates of the second- and third-order derivatives
of the energy functional that are needed to predict vibrational spectra.
Over the past decade, many techniques have been developed to efficiently 
obtain non-resonant spectra of insulating crystals, such as linear 
response theory \cite{Baroni2001, Gonze1997} for infrared absorption, 
second-order response of the density matrix \cite{Lazzeri_2003}
and the $2n+1$ theorem \cite{Veithen2005} for Raman couplings.
While being elegant and attractive, these methods 
come at the cost of rather complex code implementations,
limiting in practice applications to selected semi-local functionals
and often only to norm-conserving pseudopotentials.
This is especially the case for the third-order derivatives of the total energy,
required for the Raman calculations.
One common expedient for the latter is to resort to finite differences 
of the dielectric tensor upon atomic displacements, computed using density-functional
perturbation theory (DFPT) \cite{Baroni1986, Liang2019}.
However, this approach scales in general as $6N_{at}$ with the number
of atoms $N_{at}$, a burden in low-symmetry systems,
and becomes out of reach for the non-linear optical susceptibility tensor, 
which has been shown to be fundamental to achieve good agreement both 
in single crystals \cite{Calzolari_2013} and powder spectra \cite{Popov2020}.
An alternative framework for carrying out such derivatives is the finite numerical differentiation of forces and polarization upon the application of a homogeneous electric field\cite{Souza2002, Umari2002, Umari2003, Veithen2005, Umari2005}.
By this technique, along with finite displacements \cite{Togo2015}, one can compute straightforwardly
all the quantities needed for IR and Raman spectra, as well as allowing to use any functional (Hubbard corrected, hybrids, ...), and pseudopotential formalism (norm-conserving, ultrasoft, PAW) with little effort. 
Furthermore, the use of finite fields overcomes the expensive computation
and poor scaling of the variational dielectric tensor approach, 
allowing for the study of complex materials
such as amorphous and defective structures \cite{Umari2005,Niu2008}.
In fact, within this scheme, one would need to perform typically
only 12 ground-state calculations
in the presence of a homogeneous electric field (see Eq. \ref{eq:mixedder} and related discussion).
While appealing, this approach has seen scarce applicability,
possibly due to the many steps and pitfalls related to the underlying algorithm,
which can be highly non-trivial even for expert users.

A key solution is to streamline all of these operations through a modern workflow manager, adhering in passing to the FAIR \cite{Wilkinson2016}  principles for data sharing, while being able of automatically submitting, parsing, and processing the outputs generated by multiple calculations. 
In this study, we present a comprehensive formulation for a unified finite-displacement and finite-field approach, along with efficient Brillouin zone sampling, and encode this formalism in a self-contained Python package that operates within the AiiDA framework \cite{Huber2020, Uhrin2021}, offering a scalable computational infrastructure for automated and reproducible workflows and data provenance.

To showcase the capabilities of the approach, we demonstrate its effectiveness in predicting the infrared and Raman spectra of LiNbO$_3$ using seven different functionals: LDA, PBE, PBEsol, PBEsol+U, PBEsol+U+V, PBE0, and HSE06.

\section*{Results}

\subsection*{The finite-displacement and finite-field method}
Infrared and Raman spectroscopy are investigation methods probing 
the atomic vibrations of materials.
In both techniques, a source of light is shone on the sample,
and the incident photons will be either absorbed or scattered by the material.
In IR experiments, peaks of the absorption spectra correspond to collective 
atomic motions in resonance with that frequency, corresponding to long-wavelength polar optical phonons in crystals.
Raman spectroscopy records instead a scattered frequency $\omega_S$; in this case, the excited lattice vibration corresponds to the shift from the incoming laser frequency $\omega_L$.
Far from resonance the two phenomena are described as the change in polarization $\mathbfcal{P}$ and the change in polarizability $\chi$, for IR and Raman respectively, leading to different \textit{selection rules} for the activated phonon modes which ultimately provide a comprehensive fingerprint of the material.
The former mechanism is coupled to the Born effective charge $\mathbf{Z}_I^*$ \cite{MaxBorn1998, Baroni2001} of an atom $I$ in the unit cell, and the latter instead to the Raman
tensor defined as $d\mathbf{\chi}/d\mathbf{\tau}_I$ in the Placzek approximation \cite{Placzek1934, Brueesch1986, Porezag1996}, where $\mathbf{\chi}$ is the electronic susceptibility tensor and $\tau_I$ the atomic displacement.

In terms of these quantities, the scattering intensities of a normal mode $\nu$ are
\begin{equation}
    I_{\text{IR}}^\nu \propto \frac{|\bar{\mathbf{Z}}^{\nu} |^2}{\omega_\nu } 
    \label{eq:ir}
\end{equation}
for infrared spectra, and
\begin{equation}
    I_{\text{Raman}}^\nu 
    \propto 
    \frac{\omega_S^4}{\omega_\nu}
    ( n_{\nu}(T) + 1 )
    |\mathbf{e}_S \cdot \mathbf{\alpha}^{\nu} \cdot \mathbf{e}_L|^2
    \label{eq:raman}
\end{equation}
for Raman Stokes processes (i.e. when the scattered light frequency is lower than the incident one).
The tensors defining these amplitudes are defined via the polarization vector
\begin{equation}
    \bar{Z}^{\nu}_i  
    =
    \sum_{I,k} Z^{*}_{I,ik} 
    \frac{e^{\nu}_{I,k} }{\sqrt{m_I}}
    ~,
    \label{eq:polvec}
\end{equation}
and the Raman susceptibility tensor
\begin{equation}
    \alpha^{\nu}_{ij} 
    =
    \sqrt{\Omega}
    \sum_{I,k} 
    \frac{\partial \chi^{(1)}_{ij}}{\partial \tau_{I,k}} 
    \frac{e^{\nu}_{I,k} }{\sqrt{m_I}}
    ~,
    \label{eq:suscraman}
\end{equation}
where $\omega_\nu$ and $\mathbf{e}^{\nu}$ are the frequency and the eigenvector of the phonon $\nu$, $n_{\nu}(T)$ is the Bose-Einstein occupation function, $m_I$ the mass of the atom $I$, and $\Omega$ the volume of the unit cell
In the harmonic approximation, the phonon normal modes at zone center are obtained from the interatomic force constants (IFCs)
\begin{equation}
    \Phi_{IJ,ij}
    =
    \frac
    {
    \partial^2 E
    }
    {
    \partial \tau_{I,i} \partial \tau_{J,j}
    }
    ~,
    \label{eq:ifc}
\end{equation}
where $E$ is the potential energy surface.
The IFCs can be written in terms of first-order derivative of atomic forces $\mathbf{F}_I$ with respect to atomic displacements, and thus it can be obtained using a set of small displacement configurations, as discussed in the literature \cite{Togo2015,Baroni2001}.
The diagonalization of the mass-scaled IFCs, describing the harmonic ion dynamics, gives the phonon frequencies and eigenvectors:
\begin{equation}
    \frac
    {
    1
    }
    {
    \sqrt{m_I m_J} 
    }
    \Phi_{IJ,ij}
    e_{J,j}
    =
    \omega_{\nu}^2 e_{I,i}
    ~.
    \label{eq:eom}
\end{equation}
While the IFCs extend over an arbitrary (but decaying) number of unit cells, these photon spectroscopies 
utilize photons which carry a negligible momentum $\mathbf{q} \approx 0$, thus only probing phonons in the vicinity of the zone center, and justifying the evaluation via finite displacements in the primitive cell only.
The Born effective charges and Raman tensors can also be defined as
derivatives of forces with respect to a macroscopic electric field $\mathbfcal{E}$ \cite{Umari2005, Veithen2005}.
For the former we have:
\begin{equation}
    Z^*_{I,ij} 
    =
    \frac{\partial F_{I,j}}{\partial \mathcal{E}_i}
    ~
    \label{eq:bec}
\end{equation}
while for the latter, the Raman tensor reads:
\begin{equation}
    \frac{\partial \chi_{ij}}{\partial \tau_{I,k}} 
    =
    \frac{1}{\Omega}
    \frac{\partial^2 F_{I,k}}{\partial \mathcal{E}_i \partial \mathcal{E}_j}
    ~.
    \label{eq:ramantensor}
\end{equation}
In a finite-difference scheme, this means that a small homogeneous electric field
must be applied in order to carry out the numerical differentiation. 
In periodic-boundary conditions, this can be accomplished by extending the energy functional $E[\psi]$
to include a term \cite{Nunes2001,Souza2002,Umari2002} accounting for the coupling of the electric field and the total polarization of the system  (often referred as \textit{electric-enthalpy} functional):
\begin{equation}
    \mathcal{F}[\psi]
    =
    E[\psi]
    -
    \Omega 
    ~
    \mathbfcal{E}
    \cdot
    [
    \mathbfcal{P}^{ion}
    +
    \mathbfcal{P}^{el}[\psi]
    ]
    ~,
    \label{eq:electricenthalpy}
\end{equation}
where $\mathbfcal{P}^{ion}$ and
$\mathbfcal{P}^{el}$ are the ionic and electronic polarizations, respectively; the latter being described by the modern theory of polarization \cite{Vanderbilt1993,Resta1994}.
Such functional, while not bounded from below, admits local minima representing self-consistent stationary solutions \cite{Souza2004,Umari2004}, if $\mathbfcal{E}$ is properly chosen;
this guarantees the simultaneous evaluation of forces and electronic polarization at convergence.
As a consequence, 
the electronic polarization (that comes for free) can be exploited to extract its first and second-order derivatives with respect to an electric field \cite{Veithen2005,Umari2002}.
These correspond respectively to the high-frequency dielectric tensor $\mathbf{\epsilon}^{\infty}$, used to account for the non-analytical contribution to the IFCs \cite{Gonze1997}, and the non-linear optical susceptibility $\mathbf{\chi}^{(2)}$, key
in non-centrosymmetric crystals where the Fr\"{o}lich contribution to the Raman tensor could be sizable \cite{Calzolari_2013,Popov2020}.
Their relationship with the electronic polarization derivatives is as follows:
\begin{align}
    \epsilon^{\infty}_{ij}
    & =
    \delta_{ij}
    +
    \frac{4\pi}{\Omega}
    \frac{\partial \mathcal{P}^{el}_{i}}{\partial \mathcal{E}_{j}} 
    ~,
    \label{eq:dieltensor}
    \\
    \chi^{(2)}_{ijk}
    & =
    \frac{4\pi}{\Omega}
    \frac{\partial^2 \mathcal{P}^{el}_{i}}{\partial \mathcal{E}_{j} \partial \mathcal{E}_{k}} 
    ~.
    \label{eq:nlotensor}
\end{align}
%

\subsection*{Numerical differentiation}
The possibility of consistently describing homogenoues electric fields allows to access any derivative in  atomic displacements and electric field, while being easily extensible to advanced functionals and pseudopotential formulations \cite{Umari2005}.
We calculate here the tensors for infrared and Raman cross-sections using a central derivative formulation.
These derivatives are performed  around $\mathbfcal{E}=\mathbf{0}$,
which resembles the common experimental setup where an external
electromagnetic source is negligible, either natural or artificial.
Indicating (by dropping the indices) a computed quantity as $A$, either
a force or the electronic polarization, we calculate its $m$-th derivatives with a numerical accuracy of $n$-th order via the discretization formula:
\begin{equation}
    \frac{\partial^m A}{\partial \mathcal{E}_i^m}
    = 
    \frac{1}{\Delta \mathcal{E}^m_i}
    \sum_{l=-n/2}^{n/2} 
    ~ c_{l}^m
    ~ A 
    \left(
    l \cdot 
    \Delta \mathcal{E}_i
    \right)
    +
    \mathcal{O}(\Delta \mathcal{E}_i^n)
    ,
    \label{eq:finiteformula}
\end{equation}
where $\Delta \mathcal{E}_i > 0$ is the small electric field, which we refer to as the finite difference step, along the $i$-th Cartesian direction,
and $c_l^m$ the central derivative coefficients that can be automatically found using the Fornberg's algorithm\cite{Fornberg1988}.
The discretization is performed using uniform steps such that $l=-\frac{n}{2},\dots,0,\dots,\frac{n}{2}$ is always an integer number and the applied fields are evenly spaced (see Fig. \ref{fig:paralleldistance}(a)).
For mixed second-order derivatives of $A$, i.e. for the Raman and non-linear susceptibility tensors, we exploit a formula devised in \cite{Umari2005} to reduce the number of calculations.
We assign $ \mathcal{E}_i = \mathcal{E}_j = \lambda$ and express the mixed derivative as:
\begin{equation}
    \frac{\partial A}{\partial \mathcal{E}_i \partial \mathcal{E}_j}
    =
    \frac{1}{2}
    \left [
    \frac{\partial^2 A}{\partial \lambda^2}
    -
    \frac{\partial^2 A}{\partial \mathcal{E}_i^2}
    -
    \frac{\partial^2 A}{\partial \mathcal{E}_j^2}
    \right ]
    ~,
    \label{eq:mixedder}
\end{equation}
where each second-order derivative is evaluated through Eq. \ref{eq:finiteformula};
this can be visualized schematically in Fig. \ref{fig:paralleldistance}(a).
For example using a 2$^{nd}$ order numerical  accuracy, we would need only 12 
self-consistent field (SCF) calculations with a non-zero electric ($\Delta \mathcal{E}$-SCF) field to carry out all
the needed tensors both for IR and Raman spectra, even in low-symmetry systems
such as amorphous materials \cite{Umari2005,Niu2008};
this number can be reduced if one exploits the symmetries of the material.
In fact, there could exist symmetries belonging to the point group 
of the crystal that transform an electric field direction into an other.
For example, if the crystal satisfy inversion symmetry,
this is sufficient to reduce by a half the number of $\Delta \mathcal{E}$-SCF calculations needed, 
since forces and electronic polarization produced by $\Delta \mathcal{E}_i$
are connected by a similarity transformation to the one produced by $-\Delta \mathcal{E}_i$;
keeping the previous numerical example, we would need then just 6 $\Delta \mathcal{E}$-SCFs.
For highly-symmetric crystals, such as cubic Si, only two electric field 
directions must be evaluated; in this case, the number of
calculations reduces further to only 2.
In general, the number of $\Delta \mathcal{E}$-SCF calculations needed scales with the numerical accuracy $n$ of the central formula as
$\mathcal{N}_n = 2 \cdot \mathcal{N}_{n-2}$, where $n$ is a positive multiple of 2.

While this approach is advantageous, we also highlight that the choice of the field magnitude is restricted.
As already stated before the electric enthalpy functional \ref{eq:electricenthalpy} has no global minima,
and long-lived meta-stable states exist only below a certain critical field $\mathbfcal{E}^c$,
where the Zener tunneling is suppressed \cite{Nunes2001, Souza2002}.
An estimate of such critical value can be performed using \cite{Souza2002}:
\begin{equation}
    e |\mathbfcal{E}^c \cdot \mathbf{a}_i | 
    \simeq
    E_{\text{gap}}/N_i
    \label{eq:criticalef}
\end{equation}
where $e$ is the electric charge, $\mathbf{a}_i$ the lattice vectors defining the unitcell,
$E_{\text{gap}}$ the electronic band gap, and $N_i$ the number of k-points sampling the Brillouin zone in the $i$-th crystal direction.
For very small band gap semiconductors this critical field can be very small, 
limiting in practice the convergence due to numerical noise.
In these circumstances, a balance must be found;
nevertheless, for the majority of cases this does not represent an actual limitation.

\subsection*{Directional sampling}
A subtle issue related to the Berry phase formalism used
in the electric-enthalpy formulation is the dense sampling of
the Brillouin zone required to achieve well converged values of the
polarization, needed to carry out the numerical derivatives.
In fact, one can write the electronic polarization using the
string-averaged discretized Berry phase formulation \cite{Vanderbilt1993} 
along the direction of a primitive reciprocal lattice vector $\mathbf{b}_i$ as:
\begin{equation}
    \mathbfcal{P}_{el} \cdot \mathbf{b}_i
    =
    \frac{2 e}
    { N^{(i)}_{\perp} \Omega}
    \sum_{l=1}^{N^{(i)}_{\perp}}
    \text{Im}
    \ln
    \prod_{j=0}^{ N_{\parallel}^{(i)} -1}
    \det
    [
    S(\mathbf{k}_j^{(i)}, \mathbf{k'}_{j+1}^{(i)})
    ]
    ~.
    \label{eq:berryphasepol}
\end{equation}
Here $S_{nm}(\mathbf{k}, \mathbf{k}') = \braket{u_{n \mathbf{k}}|u_{m \mathbf{k}'}}$
represents the overlap matrix between Bloch's states $u_{l \mathbf{k}}$ belonging to occupied 
bands $l$, and $N_{\perp}^{(i)}$ is the number of strings (i.e. the number of k-points orthogonal to $\mathbf{b}_i$), each having $N_{\parallel}^{(i)}$ k-points.
Thus, when we apply a static electric field along lattice vectors,
an electronic polarization is induced according to the field direction,
properly sampled by increasing the number of k-points within each string $N_{\parallel}^{(i)}$.
While the derivatives can be performed in the crystal reference system \cite{Marzari1997},
we describe in the following some possible implementations using a Cartesian reference system (the one employed in the code).
In this coordinates system one can follow two approaches, namely either
performing an orthogonal transformation of the lattice to have
the chosen field direction along one of the new basis vectors,
or generalise Eq. \ref{eq:berryphasepol} to a generic reciprocal
vector.
It is clear it is more convenient to choose the latter,
as it requires only to modify Eq. \ref{eq:berryphasepol} to account for different $N_{\parallel}^{(i)}$ for each $N_{\perp}^{(i)}$,
while the former would require in general large supercells
to accomodate the rotated crystal.
Nevertheless, many codes use automatically generated Monkhorst-Pack (MP) meshes, 
and it is not intuitive how to properly choose a regular grid that samples correctly an arbitrary direction.
To automate the choice of such grids, we propose here the concept of
\textit{directional sampling} targeting the increase of the number of k-points in any desired direction and  crystal lattice, while still retaining a regular MP mesh.
The main idea is to define the BZ mesh by two quantities:
an \textit{orthogonal} and a \textit{parallel} k-point distance,
$d_{\perp}$ and $d_{\parallel}$ respectively.
The \textit{orthogonal} distance represents a minimum distance
between k-points in the BZ that are commonly used to achieve convergence of
total energies and forces, thus usually producing rather sparse meshes
-
this quantity is then used as a reference minimum for the generation of the grid.
The \textit{parallel} distance instead is used to define a denser number of k-points
in the desired direction, which in our case is referring to the applied electric field.
To do so, we transform the electric field direction in crystal coordinates.
This step allows us to assign weights for the sampling in reciprocal space.
Let  $\tilde{\mathbfcal{E}} = (\tilde{\mathcal{E}}_1,\tilde{\mathcal{E}}_2, \tilde{\mathcal{E}}_3)$ be the electric field in the lattice reference system, we then define the weights as:
\begin{equation}
    w_i
    =
    \frac{|\tilde{\mathcal{E}}_i |}{\max \{|\tilde{\mathbfcal{E}}| \}} 
    ~,
    \label{eq:weights}
\end{equation}
aiming at maximizing the sampling for lattice directions where the electric dipole is expected to be more relevant.
Finally, we can assign the number of k-points $N_i$ of an MP mesh as:
\begin{equation}
    N_i 
    =
    \max
    \left \{
    w_i
    \frac{|\mathbf{b}_i|}{d_{\parallel}},
    \frac{|\mathbf{b}_i|}{d_{\perp}}
    \right \}
    ~.
    \label{eq:numberofkpoints}
\end{equation}
This methodology is easy to implement and generates meshes that
better sample the direction where the induced polarization is expected,
significantly reducing the number of k-points needed instead when using 
a constant distance to define the entire MP grid.

\begin{figure}
    \centering
    \includegraphics[width=0.9\textwidth]{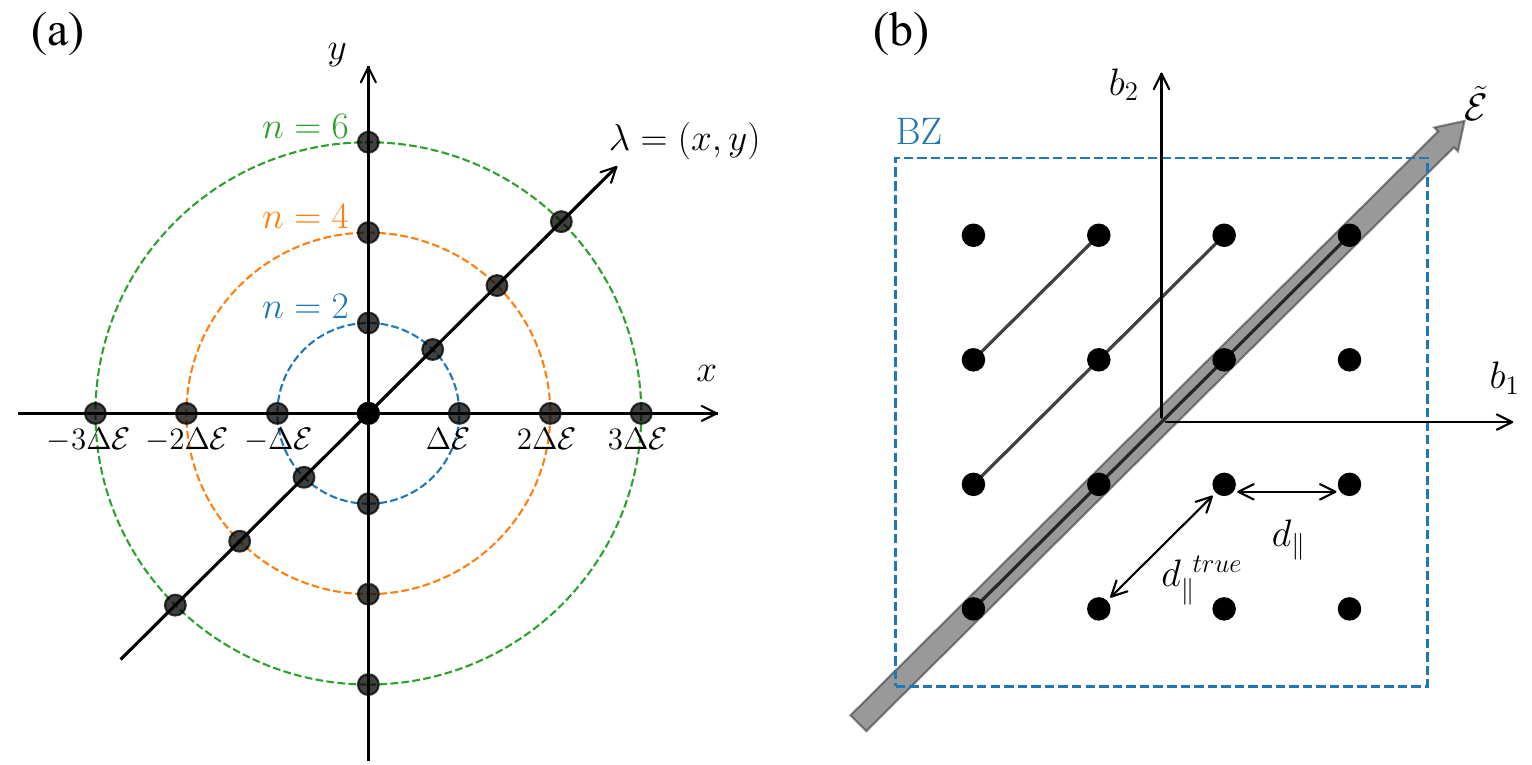}
    \caption{
    (a) Schematic illustration of the central discretization formula \ref{eq:finiteformula} and \ref{eq:mixedder} for derivatives with respect to the electric field in the directions $i=\{x,y,\lambda=(x,y)\}$. The black dots represent the applied electric fields $l \Delta\mathcal{E}_i$ used in the SCF calculations to evaluate the electric-field-dependent forces and polarization to employ in \ref{eq:finiteformula} and  \ref{eq:mixedder}, while the different coloured circles are associated to the numerical accuracy $n$ achieved in the numerical differentiation (here shown up to $n=6$). It can be noticed that the evaluation of the numerical derivative at $n>2$ provides automatically derivatives at lower accuracy $n$; this is helpful to check the convergence \textit{a posteriori} in respect to the finite step size $\Delta\mathcal{E}$. 
    (b) Square reciprocal lattice showing a shifted regular k-points sampling (black dots) of the BZ generated via Eqs. \ref{eq:weights}-\ref{eq:numberofkpoints} for the shown electric field $\tilde{\mathbfcal{E}}$ in reciprocal coordinates.
    The strings of k-points are indicated by straight black lines, and are only shown on part of the BZ for clarity.
    In this particular case $\tilde{\mathcal{E}}_1 = \tilde{\mathcal{E}}_2$, hence $w_1=w_2=1$ from Eq. \ref{eq:weights}, meaning the generated mesh via Eq. \ref{eq:numberofkpoints} have k-points along $b_1$ and $b_2$ equally spaced by $d_{\parallel}$, indicated on the figure by an horizontal arrow.
    It can be seen that this introduced distance  differs from the actual distance $d^{true}_{\parallel}$ (diagonal arrow) among string-connected k-points.
    }
    \label{fig:paralleldistance}
\end{figure}

We point out that the \textit{parallel} distance defined
may not resemble the \textit{true} distance among k-points within strings
along a defined direction.
This is easily understood with a trivial example reported in Fig. \ref{fig:paralleldistance},
which represents a simple two-dimensional BZ lattice sampling with
an applied field along a mixed direction.
It is clear that the distance among adjacent k-points within strings 
will have a greater distance $d^{true}_{\parallel}$ than the $d_{\parallel}$ introduced
due to geometrical reasons.
It is understood that our simplistic solution might be in the future substituted with generalised MP grids \cite{Wang2021} that can be \textit{ad hoc} fine tuned for effective and inexpensive samplings.
However, this is far from being a trivial task and thus out of scope for the present work.

\subsection*{Computational workflow}
\begin{figure}
    \centering
    \includegraphics[width=1.0\textwidth]{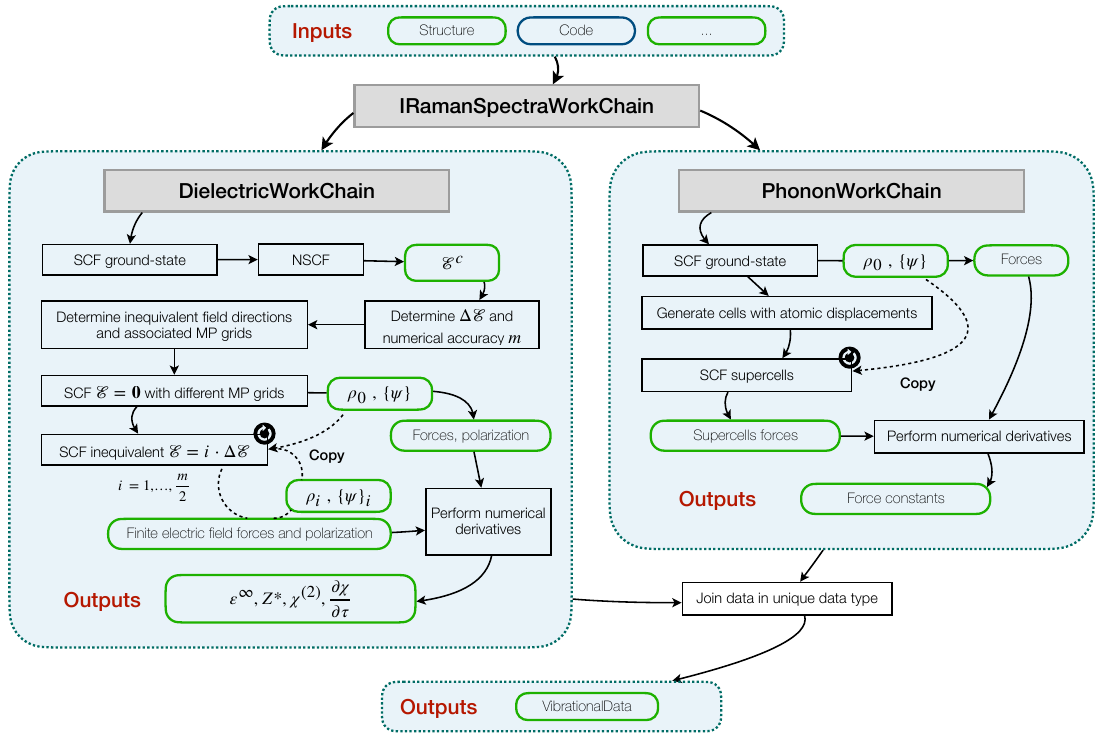}
    \caption{
    Computational workflow scheme of the \texttt{IRamanSpectraWorckChain} and of its sub-workflows, the \texttt{DielectricWorkChain} and the \texttt{PhononWorkChain}. The inputs can be easily, fully defined via the \texttt{get\_builder\_from\_protocol} while specifying only a \texttt{StructureData} and a \texttt{Code}.
    The black circles on top of boxes mean that the calculation starts using charge density and wavefunctions from a previous calculation.
    }
    \label{fig:workflows}
\end{figure}

In this section, we finally present the computational workflows
that allow to simulate the desired spectra, which can be infrared, Raman, or both, in an automated fashion.
We designed two main independent workflows as AiiDA \texttt{WorkChains} \cite{Huber2020}: one carrying out solely the finite displacement
calculations for the phonon modes to evaluate the IFCs, which we called \texttt{PhononWorkChain},
and the other for the mixed total-energy derivatives via the homogeneous electric fields, the \texttt{DielectricWorkChain}.
When joined together, the above two provide the \texttt{IRamanSpectraWorkChain}, as schematically outlined  in Fig. \ref{fig:workflows}.
%

The \texttt{IRamanSpectraWorkChain} takes as inputs 
a structure, a code and other details on how to run the workflows (e.g. SCF parameters, wall-time, parallelization options, and so on).
These information are then split and/or shared between the two sub-workflows.
Both start running an initial SCF ground-state calculation,
meant to produce the charge density $\rho_0$ and wavefunctions $\{\psi_i\}$ that will be used as a starting point for the next runs.
We generally represent this restart throughout the flowchart of Fig. \ref{fig:workflows} by a curved white arrow in a black circle, and we use the dashed arrows to indicate that the files have been copied over to a fresh new folder. 
When these are finished, the \texttt{PhononWorkChain} generates the  structures with irreducible
atomic displacements 
and computes their forces.
The latter are then used to produce the matrix of IFCs,
which will be employed in a post-processing step to obtain the phonon  frequencies and eigenvectors.
In parallel, the \texttt{DielectricWorkChain} evaluates $\mathcal{E}^c$
through a non-SCF calculation and extracts the numerical accuracy $n$ and the electric field step $\Delta \mathcal{E}$, as follows:
when the critical value is larger than $10^{-4}~\text{(Ry)a.u.}$
($Rydberg$ atomic units;  1 (Ry)a.u. $\approx 36.3609$ V/\AA ), a 4$^{th}$ order numerical expansion is chosen, otherwise a 2$^{nd}$ order is used.
The 4$^{th}$ order allows to achieve a greater numerical accuracy for the finite differences, thus it is utilized whenever it is possible to apply an electric field above threshold.
In cases the electric field that can be applied is smaller than $10^{-4}~ \text{(Ry)a.u.}$, a 2$^{nd}$ order expansion is sufficient, thus avoiding calculations that would be affected by too much numerical noise \cite{Porezag1996}; the amplitude of the electric field step in fact may result too small to produce significant digit changes in forces and polarization, hence possibly worsening the numerical derivative even if adding extra points to the formula.
The step $\Delta \mathcal{E}$ is instead selected according to:
\begin{equation}
    \Delta \mathcal{E}
    =
    \begin{cases}
    \frac{10^{-3}}{n} & \text{if $\mathcal{E}^c > 10^{-3}$} , \\
    \frac{ \text{round($\mathcal{E}^c$, 5)}}{n} & \text{if $10^{-4} < \mathcal{E}^c \leq 10^{-3}$} , \\
    \frac{ \text{round($\mathcal{E}^c$, 6)}}{n} & \text{if $\mathcal{E}^c \leq 10^{-4}$} ,
    \end{cases}
    \label{eq:stepselection}
\end{equation}
where the electric field units are again expressed in (Ry)a.u. and \textit{round(x,y)} rounds the number $x$ to the $y$ digit after the decimal point.
These choices should guarantee larger values of the finite step to reduce the numerical noise on forces and polarization, and at the same time higher accuracy to remove the step size dependence of the finite difference formula, which can be verified a posteriori when $n>2$ (see also Fig. \ref{fig:paralleldistance}(a)).
We point out here that eventually the user can still enforce the accuracy $n$ and/or the $\Delta\mathcal{E}$ value that will be used in Eqs. \ref{eq:finiteformula} and \ref{eq:stepselection}. 
%
Once an electric field is chosen and the ground-state SCF calculation is finished, a symmetry analysis is performed to find the inequivalent directions $\hat{\mathbfcal{E}}$ of the electric fields to apply, where $\hat{\mathbfcal{E}}$ is a versor.
For each such inequivalent direction an MP mesh is determined through Eq. \ref{eq:numberofkpoints} and a series of SCF calculations are then launched at $\mathbfcal{E}=\mathbf{0}$ with the corresponding mesh restarting from the ground-state density of the previous SCF.
The reason for computing the SCF at $\mathbfcal{E}=\mathbf{0}$ with these meshes is twofold: first, it is crucial to evaluate the polarization operator, set up in following finite electric-field calculations at the corresponding mesh reference, and, second, it guarantees the same level of convergence on forces and polarization for the finite differentiation.
In fact, $\hat{\mathbfcal{E}}$ and $-\hat{\mathbfcal{E}}$ produce the same k-point meshes (as can be verified via Eqs. \ref{eq:weights}-\ref{eq:numberofkpoints}), hence all the calculated values, at ${\mathbfcal{E}} = \mathbf{0}$ included,  employed in the finite formula \ref{eq:finiteformula} have the same level of convergence with respect to the particular k-point grid.
Once finished, electric-enthalpy calculations start for each inequivalent direction $\hat{\mathbfcal{E}}$ at the smallest step value $\Delta \mathbfcal{E} = \Delta \mathcal{E} \cdot \hat{\mathbfcal{E}}$.
If the numerical accuracy $n$ is greater than 2, further SCF calculations with
homogeneous electric fields are run, restarting from the previous steps till the maximum value $(n/2)\Delta \mathbfcal{E}$; this loop corresponds to increasing/decresaing the value of $l$ in Eq. \ref{eq:finiteformula}.
At this point, the computed forces and polarizations are gathered together to perform the numerical differentiation according to Eqs. \ref{eq:finiteformula} and \ref{eq:mixedder} for achieving second and third rank tensors \ref{eq:bec}-\ref{eq:ramantensor}-\ref{eq:dieltensor}.
As a final step, the \texttt{IRamanSpectraWorkChain} joins all
the information in a unique AiiDA data type,
named \texttt{VibrationalData}, able to efficiently store
large arrays and information in the AiiDA repository and database, and with provenance graph for full reproducibility \cite{Huber2020}.
This data type is a self-contained Python class, and it is provided
with a number of post-processing methods and features that allow the user 
to compute powder and single crystal polarized vibrational intensities and frequencies,
along with the symmetry labels of the respective phonon modes.
We also implemented an \texttt{IntensitiesAverageWorkChain}
which performs the spherical average spectra, important in Raman for
non-centrosymmetric crystals \cite{Popov2020}.
%

As mentioned earlier, the inputs of these workflows can be fully customized by the user.
We do not report here the full list of these inputs,
which can be explored through the usual command line interface of AiiDA \cite{Huber2020}
and through the on-line documentation made available.
However, we would like to emphasize the possibility of fully defining
the entire set of inputs via a limited set of them,
as already been done in other AiiDA packages \cite{Uhrin2021}.
In similar fashion to Ref. \cite{Uhrin2021}, we identified the fundamental inputs
to be the \texttt{code}, i.e. the Quantum ESPRESSO \cite{Giannozzi2009,Giannozzi2017, Giannozzi2020} \textit{pw.x} binary, associated with the computer, that runs the calculations, the relaxed \texttt{structure} and the \texttt{protocol}, a string defining
in a general and user-friendly fashion the accuracy of the simulation.
This minimal set, used through the \texttt{get\_builder\_from\_protocol} method \cite{Uhrin2021} designed for all the workchains, provides the user with a pre-filled \texttt{builder} (the main unit in AiiDA used to run workflows) ready to be submitted.
This is adequate for non experts in the field that might be interested in simulating their structure's spectra with minimal knowledge, but also in the context of high-throughput searches or for teaching or testing.

\subsection*{All-functionals LiNbO$_3$ spectra}
We demonstrate the power of this approach studying the ferroelectric phase of LiNbO$_3$ (space group R3c, no. 161) for which the vibrational spectra have been long debated in the literature \cite{Veithen2002, Margueron2012, Sanna2015, Kojima2016} and exhibit large LO-TO phonon splitting at $\Gamma$ \cite{Veithen2002} that are ideal for exploring different functional flavours.
We selected three commonly used local and semi-local funcationals, LDA, PBE and PBEsol, then Hubbard-corrected PBEsol in the standard and extended version, namely DFT+U \cite{Himmetoglu2011, Himmetoglu2013} and DFT+U+V \cite{Jr2010, Cococcioni2019, Timrov2020}, and two of the most used hybrids functionals for inorganic crystalline materials, HSE06 \cite{Heyd2003, Heyd2006} and PBE0 \cite{Perdew1996}.

\begin{figure}
    \centering
    \includegraphics[width=\textwidth]{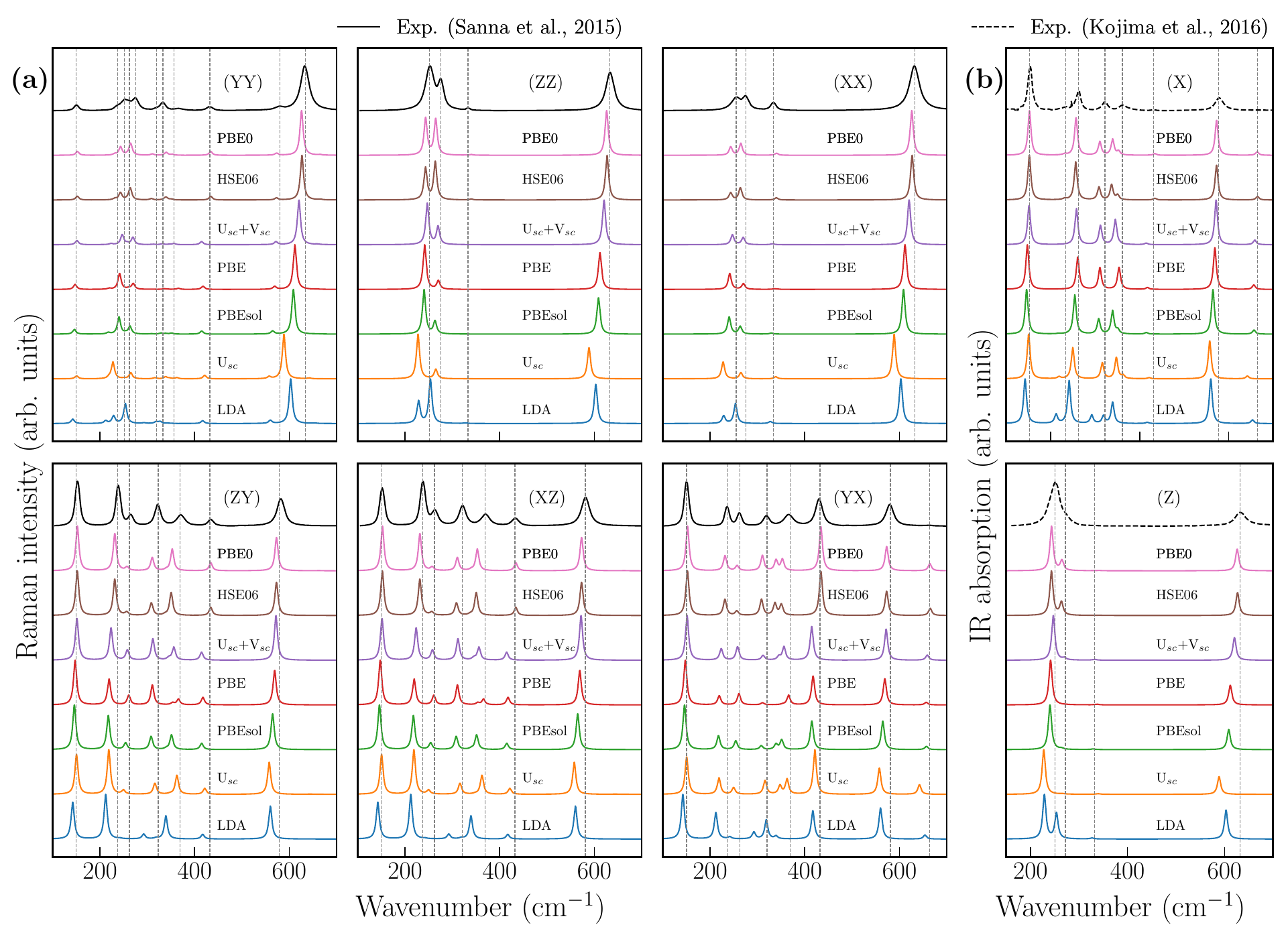}
    \caption{Raman and IR spectra of ferroelectric LiNbO$_3$ as obtained with seven different functional, and compared to experimental measurements. (a) Raman single-crystals spectra for different polarization setups (reported in paranthesis) for pure TO modes corresponding to E and A$_1$ symmetries. (b) Infrared absorption using polarized light for E (upper panel) and A$_1$ (lower panel) symmetry modes. The theoretical spectra are ordered according to the error $\overline{|\Delta \omega|}$ (shown in Tab. \ref{tab:statistics}), from smallest (pink, PBEsol+U$_{sc}$+V$_{sc}$) to largest (blue, LDA). Hubbard-corrected PBEsol$+U_{sc}$ and PBEsol$+U_{sc}+V_{sc}$ are labeled just $U_{sc}$ and $U_{sc}+V_{sc}$, the subscript \textit{sc} refers to the self-consistent calculation of $U$ and $V$ \cite{Timrov2018, Timrov2021}. Theoretical intensities are smeared with a constant 8 cm$^{-1}$ broadened Lorentzian.
    }
    \label{fig:linb03spectra}
\end{figure}

\begin{figure}
    \centering
    \includegraphics[width=0.3\textwidth]{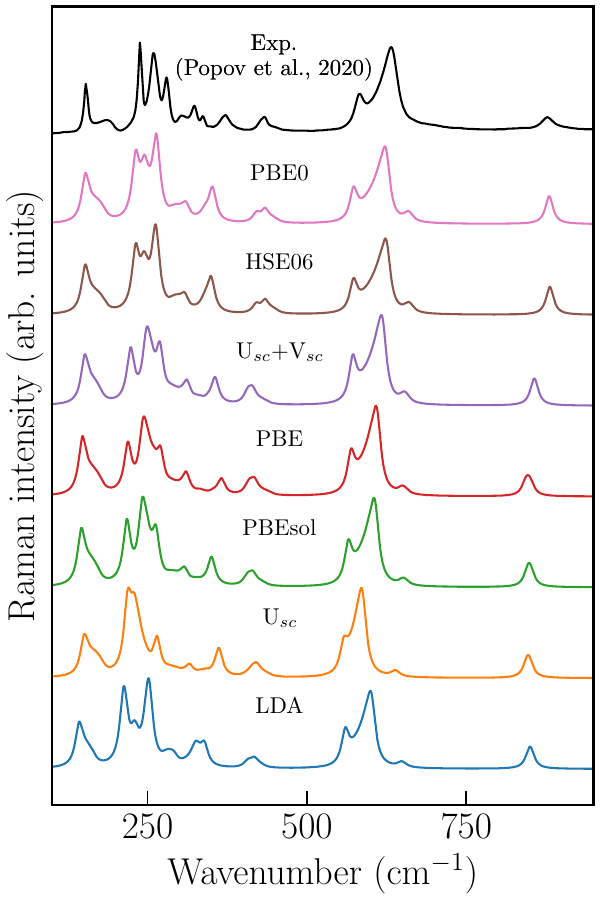}
    \caption{
    Raman powder spectra using the spherical average formula \cite{Popov2020} and accounting for the Fr\"{o}lich term to reproduce the characteristic asymmetric line shapes.
    Theoretical intensities are smeared with a constant 14 cm$^{-1}$ broadened Lorentzian.
    The theoretical spectra are reported in order of decreasing error $\overline{|\Delta \omega|}$, and Hubbard-corrected PBEsol are abbreviated only with their corrective parameters, as in Fig. \ref{fig:linb03spectra}.
    }
    \label{fig:powderraman}
\end{figure}

Through the use of the \texttt{IRamanSpectraWorkChain}, we carried out the simulation of  the vibrational spectra for all the functionals  above and we compare our results to experimental measurements from the most recent literature \cite{Popov2020, Sanna2015, Kojima2016}.
In particular, we compare our simulations in Fig. \ref{fig:linb03spectra} to (a) different Raman single-crystal polarization setups \cite{Sanna2015} and (b) two polarized infrared absorption spectra \cite{Kojima2016}, and
in Fig. \ref{fig:powderraman} to polycristalline Raman powder spectra \cite{Popov2020} for which we use the spherical average procedure accounting for the $\chi^{(2)}$ contribution, as outlined in Ref. \cite{Popov2020}.
The single-crystal data allowed us to compare directly the frequencies and, in the case of Raman, the intensities.
In (a) and (b), we ranked the spectra in the plots in function of the statistical estimator $\overline{|\Delta \omega|} = \sum_{\nu=1}^N|{\omega_{\nu}^{\text{exp}}-\omega_{\nu}^{\text{theo}}}|/N$, where $N$ is the number of active modes detected.
%
%
%
All the functionals show good agreement with experiments and with the results of the literature \cite{Veithen2002,Sanna2015}.
PBEsol+U$_{sc}$+V$_{sc}$, PBE, and the two hybrids show competitive results among each other,
while PBEsol, PBEsol+U$_{sc}$ and LDA have quantitative larger discrepancies.
In particular in Tab. \ref{tab:statistics} we report statistical estimators for the comparison of the theoretical and experimental phonon frequencies and intensities for Fig. \ref{fig:linb03spectra}(a).
%
\begin{table}
    \centering
    \begin{tabular}{lccccccc}
        \toprule
        {} & LDA & PBEsol+U$_{sc}$  & PBEsol & PBE & PBEsol+U$_{sc}$+V$_{sc}$ & HSE06 & PBE0  \\
        \hline
        $\overline{|\Delta \omega|}$ & 21.3 & 16.5 & 14.4 & 10.8 & 9.7 & 9.3 & 8.9 \\ 
        $\overline{\Delta \omega}$ & 20.8 & 13.2 & 13.0 & 7.7 & 7.5 & 5.8 & 5.3 \\ 
        $|\Delta \omega|_{max}$ & 37.2 & 33.8 & 27.3 & 22.0 & 21.6 & 24.0 & 23.0 \\ 
        $\overline{|\Delta \omega|^*}$ & 13.8 & 13.0 & 8.8 & 9.2 & 8.7 & 9.8 & 9.6 \\ 
        $\overline{|\Delta I|}$ & 37 & 22 & 19 & 21 & 15 & 27 & 25 \\ 
        \toprule
    \end{tabular}
    \caption{
    Statistical error estimators between functional predictions and experiments, given in cm$^{-1}$ for frequencies $\omega$ and in percentage for the intensities $I$.
    The estimators are:
    $\overline{|\Delta \omega|} = \sum_{\nu=1}^N|{ \Delta \omega_{\nu} }|/N$ is the mean absolute error;
    $\overline{\Delta \omega} = \sum_{\nu=1}^N{ \Delta \omega_{\nu}  }/N$ is the mean error;
    $|\Delta \omega_{max}|$ is the absolute max difference;
    $\overline{|\Delta \omega|^*} =
    \sum_{\nu=1}^N | { \Delta \omega_{\nu} 
    -\overline{\Delta \omega} } | / N $ 
    is the absolute error for frequencies shifted by their mean difference; 
    $\overline{|\Delta I|} = 100 \cdot \min_f \{ \Delta I(f) \}$ represents the percentage error of the optimal scaled intensities;
    $N$ is the number of considered modes and $\Delta \omega_{\nu}$ is the difference between experimental and theoretical functional  mode frequency.
    }
    \label{tab:statistics}
\end{table}
%
These spectra setups allowed to accurately estimate the error of the intensities evaluated through the minimization of the relative error:
\begin{equation}
    \Delta I (f) = \sqrt{
        \frac{
        \sum_{\nu=1}^N ( {I_{\nu}^{\text{theo}} - f I_{\nu}^{\text{exp}}} )^2
        }
        {
        \sum_{\nu=1}^N ( f I_{\nu}^{\text{exp}} )^2
        }
    }
    \label{eq:delta}
\end{equation}
where $f$ represents the unknown scaling factor between theoretical $I^{theo}$ and experimental $I^{exp}$ intensities, due to the use of arbitrary units in the experimental spectra.
The smallest absolute error on frequencies $\overline{|\Delta \omega|}$ is provided by the PBE0 functional with 8.9 cm$^{-1}$, and on intensities $\overline{|\Delta I|}$ by the PBEsol+U$_{sc}$+V$_{sc}$ functional with 15\%.
It is to note that the latter also shows a remarkable $\overline{|\Delta \omega|}$ of 9.7 cm$^{-1}$, less than 1 cm$^{-1}$ apart from the best performing functional for this material, whereas the former has 25\% error for $\overline{|\Delta I|}$, 10\% more with respect to PBEsol+U$_{sc}$+V$_{sc}$.
We also reported the renormalized absolute error $\overline{|\Delta \omega|}^*$, for which the frequencies are shifted by the bare mean error $\overline{\Delta \omega}$ before the modulus average; this gives some insights whether the discrepancy arises from a systematic error, e.g. due to the geometry used.
Interestingly, after renormalization, most of the functionals get rather close error values.
While this might be due to the renormalization of the phonon frequencies due to the thermal expansion which tends to soften the vibrational modes, all the theoretical spectra were performed at constrained experimental lattice (see the Method section), meaning the systematic error is unrelated to this effect. 
While other thermal contributions should in principle be accounted for, 
their inclusion would require more sophisticated level of theory \cite{Monacelli2018, Monacelli2021, Monacelli2021a} which goes beyond the scope of the present work.
Moreover, at the experimental conditions of our references, such contributions are expected to produce only minor phonon shifts.
Given these points we might conclude in this case that such errors are intrinsic to the functionals.
%

\section*{Discussion}
In this paper, we have presented a comprehensive and efficient approach that utilizes finite displacements and finite fields to calculate infrared and Raman spectra with any modern exchange-correlation functional and pseudopotential formulation (norm-conserving, ultrasoft, PAW).
Our approach overcomes the current expensive alternative of the variational dielectric tensor technique \cite{Liang2019} for Raman coupling calculations, which remains impractical in large or low symmetry systems.
This is also very relevant when generating coupling tensors in low-symmetric supercells to incorporate temperature and quantum effects \cite{Monacelli2021,Monacelli2021a,Siciliano2023}, or to train a tensorial machine-learning potential useful for molecular dynamics simulations.

The formalism has been implemented as a highly optimized, automatic workflow package
for the prediction of vibrational spectra.
The technical and practical challenges have been successfully addressed with the help of the AiiDA infrastructure \cite{Huber2020},
which empowered the design of fully reproducible, reusable, user-friendly workflows for one of the most important class of spectroscopic techniques for materials characterisation.
As such, we believe it will be broadly useful to both the computational community and the experimental community at large.

Furthermore, we have shown that the workflows can effectively operate with state-of-the-art functionals and can be employed for a comprehensive analysis of both single crystals and powder spectra.
In conclusion, we believe that the present approach will pave the way for accelerated materials characterization of materials displaying complex, challenging chemistry.
%

\section*{Methods}
%
The infrared and Raman calculations were carried out using the workflows developed, relying on aiida-phonopy \cite{Bastonero2022} for pre- and post-processing and aiida-quantumespresso \cite{Huber2020, Uhrin2021}, having Quantum ESPRESSO \cite{Giannozzi2009, Giannozzi2017, Giannozzi2020} as DFT quantum engine.
Before performing the vibrational spectra calculations, the positions of the atoms in the LiNbO$_3$ primitive cell are optimized at fixed experimental lattice geometry \cite{Weigel2020} for all functionals until the total energy and forces acting on atoms are less than $10^{-4} ~ \text{Ry/atom}$ and $5 \cdot 10^{-5} ~ \text{Ry/Bohr}$, respectively. 
Norm-conserving pseudopotentials from the Pesudo-Dojo library \cite{Setten2018} are used for the LDA and hybrid functionals, employing PBE pseudopotentials for the latter, and pseudopotentials from the precision SSSP (version 1.1) \cite{Prandini2018, Garrity2014, Schlipf2015, Willand2013, Corso2014, Topsakal2014, Setten2018} library for the others.
A cutoff of 90 Ry is used for the wavefunctions; 4 and 12 times greater cutoffs are used for the charge density (respectively for Pseudo-Dojo and SSSP based pseudopotentials).
The Brillouin zone is sampled uniformly with a 0.3 \AA$^{-1}$~ k-points distance.
The Hubbard parameters are computed following the self-consistent
procedure exploiting density-functional perturbation theory,
as explained in \cite{Timrov2018, Timrov2021, Timrov2022}, using a q-point distance of 0.4 \AA$^{-1}$, orthogonalized atomic orbitals \cite{Floris2020, Timrov2021} for the occupation matrices, and with onsite (intersite) parameters converged within 0.01 (0.005) eV (a very conservative threshold).
The hybrid calculations were performed using a 90 Ry cutoff and a single q-point mesh for the exact exchange operator, and the adaptively compressed exchange (ACE) technique \cite{Lin2016} was used to speed up the convergence of the calculations.
Finally, finite differences were carried out using a 0.01 \AA~ displacement distance and a denser k-point distance of 0.15 \AA$^{-1}$~ for the phonon calculations (in the \texttt{PhononWorkChain}), and an electric field step of $5 \cdot 10^{-4}$ (Ry)a.u., a 2$^{nd}$-order numerical accuracy, and a k-point distance of 0.15 \AA$^{-1}$~ in the parallel direction of the applied electric field for the electric-field derivatives (as carried out in the \texttt{DielectricWorkChain}).

\bibliography{main}


\section*{Data availability}

The data produced in this work can be found on the Materials Cloud Archive\cite{Bastonero2023}.

\section*{Code availability}

The source code is made available open-source on GitHub (https://github.com/bastonero/aiida-vibroscopy). It is also distributed as an installable package through the Python Package Index (https://pypi.org/project/aiida-vibroscopy/).
\section*{Acknowledgements}

L.B. would like to thank (in alphabetic order) Marnik Bercx, Chiara Cignarella, Virginie de Mestral, Michele Kotiuga, Eric Macke, Giovanni Pizzi,  Norma Rivano and Iurii Timrov for fruitful discussions.
The authors gratefully acknowledge support from the Deutsche Forschungsgemeinschaft (DFG) under Germany’s Excellence Strategy (EXC 2077, No. 390741603, University Allowance, University of Bremen) and Lucio Colombi Ciacchi, the host of the “U Bremen Excellence Chair Program,”
as well as computing time granted by the Resource Allocation Board and provided on the supercomputer Lise and Emmy at NHR@ZIB and NHR@G\"{o}ttingen as part of the NHR infrastructure. 
The calculations for this research were conducted with computing resources 
under the project \texttt{hbc00053} and \texttt{hbi00059}.
\\

\section*{Author contributions}

L.B. designed and developed the code, wrote the first draft of the manuscript, ran and analysed the calculations.
N.M. supervised the project, discussed the results, and contributed to the final version of the manuscript.

\section*{Competing interests}
The authors declare no competing interests.

\section*{Additional information}

Correspondence should be addressed to L.B. 

\end{document}